\documentclass[pra,twocolumn,preprintnumbers,showpacs,amsmath,amssymb]{revtex4}
\usepackage{amsfonts}
\usepackage{tipa}

\usepackage{epsfig,graphicx}
\usepackage{amstext}
\usepackage{amsmath}
\usepackage{graphicx}

\begin{document}


\title{Effects of impurity on fidelity of quantum state transfer
       via spin channels}

\author{Wen-Wen Zhang$^1$, Ming-Liang Hu$^{1,}$\footnote{Corresponding author.\\
        {\it E-mail address:} mingliang0301@xupt.edu.cn (M.-L. Hu).},
        Dong-Ping Tian$^{1,2}$}
\address{$^1$School of Science, Xi'an University of Posts and
             Telecommunications, Xi'an 710061, China
       \\$^2$Xi'an University of Architecture and Technology,
             Xi'an 710055, China}

\begin{abstract}
   By adopting the concept of fidelity, we investigated efficiency of
quantum state transfer with the XX chain as the quantum channel.
Different from the previous works, we concentrated on effects of
spin and magnetic impurity on fidelity of quantum state transfer.
Our results revealed that the spin impurity cannot prevent one from
implementing perfect transfer of an arbitrary one-qubit pure state
across the spin channel, however, the presence of magnetic impurity
or both spin and magnetic impurities may destroy the otherwise
perfect spin channels.
\end{abstract}


\maketitle
\section{Introduction}
   Since the seminal work of Bose \cite{R1}, quantum state transfer
along linear arrays of interacting qubits or spins \cite{R2,R3},
which is closely associated with its potential applications in
quantum communication \cite{R4,R5,R6,R7}, has been discussed by a
number of authors
\cite{R8,R9,R10,R11,R12,R13,R14,R15,R16,R17,R18,R19, R20,R21}. In
Bose's scheme \cite{R1}, the state to be transmitted is initially
encoded at one end of an unmodulated XXX spin chain by the sender
Alice, then time evolution of the system, and after certain
intervals of time, the state will be received by the receiver Bob at
another end of the chain with some fidelity. Since then, many
schemes focused on the implementation of perfect quantum state
transfer (i.e., transferring a quantum state with fidelity equals to
unity) by adopting pre-engineered spin chains as quantum channels
have been proposed \cite{R11,R12,R13,R14,R15,R16,R17, R18,R19,R20}.
In particular, by using the identity of an $N$-qubit XX spin chain
with a fictitious spin-${1\over 2}(N-1)$ particle, Christandl et al.
\cite{R11} found that if the nearest-neighbor (NN) coupling strength
$J_i$ ($i$ is the spin index) satisfying the equality
${J_i}={J_{N-i}}=\lambda\sqrt{i(N-i)}$ ($\lambda$ is a scaling
constant), then perfect state transfer across this modulated spin
chain can always be achieved. This finding has been experimentally
tested by Zhang and Long et al. in a recent work \cite{R21} by using
liquid nuclear magnetic resonance (NMR) system. Moreover, taking
advantage of the so-called Jordan-Wigner transformation, Karbach and
Stolze \cite{R13} outlined a general procedure for designing spin
chains with NN interactions for perfect quantum state transfer. Then
Kay demonstrated that perfect state transfer is also possible in the
presence of next-nearest-neighbor (NNN) couplings \cite{R18}. In a
more recent work \cite{R19}, Kostak et al. studied the issue of
perfect state transfer in networks of arbitrary topology and
coupling configuration. Significantly, all the previous results
about perfect state transfer can be understood within the
theoretical framework established in Ref. \cite{R19}. Except the
above-mentioned works, effects of decoherence caused by surrounding
environment on fidelity of quantum state transfer across a spin
chain has also been discussed recently \cite{R22,R23,R24}.

   Here we investigate efficiency of quantum spin channel with the
presence of impurities. The idea is stimulated by the fact that in
real experiments, different kinds of impurities or imperfections are
likely to be present in the solid-state systems. Impurity plays an
important role in condensed matter physics, and its effects on
properties of various quantum systems have been discussed by a
number of authors \cite{R25,R26,R27} historically. Recently,
research interest in this issue has been revived due to the
important role impurity plays in quantum information processing
(QIP) tasks \cite{R28,R29,R30,R31,R32,R33,R34}. Particularly, the
research results showed that sometimes even a small defect may
destroy the entanglement of the system completely, while in some
other cases the impurity can also be used to enhance amount of
entanglement \cite{R32,R33,R34}. This fact naturally arises the
following question: how the otherwise perfect quantum spin channel
works if impurities are introduced? The purpose of this paper is to
address this issue by investigating average fidelity of quantum
state transfer across an XX spin chain with the presence of spin
impurity, magnetic impurity \cite{R29} as well as both spin and
magnetic impurities. Our results revealed that the presence of spin
impurity cannot rule out the possibility of perfect state transfer,
while the presence of magnetic impurity or both spin and magnetic
impurities may destroy the otherwise perfect spin channels.

   The structure of this paper is arranged as follows. In Section
II, we examined effects of spin impurity on average fidelity of
quantum state transfer by using the XX chain as the quantum channel,
and gave the corresponding methods to maximize the average fidelity
to its maximum value 1. Then in Sections III and IV, the calculation
in the preceding section is repeated by changing the spin impurity
to magnetic impurity as well as both spin and magnetic impurities,
respectively, through which we show that the otherwise perfect
quantum spin channel will be destroyed for these two cases. We also
demonstrated how to minimize the detrimental effects introduced by
magnetic impurity as well as both spin and magnetic impurities by
performing local unitary operations in these two sections. Finally,
we concluded this paper in Section V.

\section{The quantum channel with spin impurity}

   In this section, we examine effects of spin impurity (denoted by
a spin-1 particle) on state transfer in an XX spin chain. We assume
the quantum state to be transmitted is encoded at the first spin as
$|\varphi_{in}\rangle=\cos{(\theta/2)}|0\rangle+e^{i\phi}
\sin{(\theta/2)}|1\rangle$ (with $|0\rangle$ and $|1\rangle$
represent the state of spin up and down, respectively), and all the
other spins in the chain are initialized to the ground state
$|0\rangle$, thus the initial state of the whole system at time
$t=0$ becomes
\begin{equation}
 |\psi(0)\rangle=\cos{\theta\over 2}|\textbf{0}\rangle+e^{i\phi}
                 \sin{\theta\over 2}|1\rangle,
\end{equation}
where $|\textbf{0}\rangle=|0_1,0_2,\ldots,0_N\rangle$ and
$|1\rangle=|1_1,0_2,\ldots,0_N\rangle$, with $\theta\in[0,\pi]$ and
$\phi\in[0,2\pi]$ being the polar and the azimuthal angles,
respectively.

   We first consider efficiency of quantum state transfer by using the
two-spin XX chain as the quantum channel, with the impurity spin
locating at the first site. Then the Hamiltonian of the system can
be expressed as
\begin{eqnarray}
\hat H=J(S_1^x s_2^x+S_1^y s_2^y)+B(S_1^z+s_2^z),
\end{eqnarray}
where $s_i^\alpha$ and $S_i^\alpha$ $(\alpha=x,y,z)$ denote the
spin-1/2 and spin-1 operators (in units of $\hbar$) at the $i$th
site (same notations are used throughout this paper). $J$ and $B$
are the coupling strength between the two neighboring spins and the
intensity of the external magnetic fields applied to the two-spin
system.

   For the initial state $|\psi(0)\rangle$, the state at a given
time, say $t$, is represented by $|\psi(t)\rangle=e^{-i\hat
Ht}|\psi(0)\rangle$. From the explicit form of the system
Hamiltonian $\hat H$, one can show that its dynamics is completely
determined by the time evolution in the zero and single excitation
subspace $\mathcal{H}_{0\oplus 1}$, thus it suffices to restrict our
attention to the dynamics of
$\rho(t)=|\psi(t)\rangle\langle\psi(t)|$ in this 3-dimensional
subspace spanned by $\{|\textbf{0}\rangle, |1\rangle, |2\rangle\}$,
which yields
\begin{eqnarray}
 |\psi(t)\rangle=\cos{\theta\over 2}f_0|\textbf{0}\rangle+
 e^{i\phi}\sin{\theta\over 2}\sum_{n=1}^2 f_n|n\rangle,
\end{eqnarray}
where
$|n\rangle=|0_1,0_2,\ldots,0_{n-1},1_n,0_{n+1},\ldots,0_N\rangle$
denotes the site basis ($N$ is the number of sites in the chain,
here $N=2$), and the other three parameters $f_n$ ($n=0,1,2$) are
given by
\begin{eqnarray}
 f_0=\langle\textbf{0}|e^{-i\hat Ht}|\textbf{0}\rangle,\,
 f_n=\langle n|e^{-i\hat Ht}|1\rangle.
\end{eqnarray}

   In the present paper, we adopt the concept of average fidelity
(the fidelity $F=\langle
\varphi_{in}|\rho_{2}(t)|\varphi_{in}\rangle$ averaged over all pure
input states in the Bloch sphere) $\bar F={1\over 4\pi}\int F
d\Omega={1\over 4\pi}\int_0^{2\pi}d\phi\int_0^\pi d\theta
F\sin\theta$ as an estimation of the quality of state transfer from
the sender Alice to the receiver Bob \cite{R1}. For state
$|\psi(t)\rangle$, the reduced densit matrix $\rho_2(t)$ can be
obtained by tracing qutrit 1 from $\rho(t)$ as
\begin{equation}
 \rho_{2}(t)=\left(\begin{array}{cc}
  1-\sin^2\frac{\theta}{2}|f_2|^2 & {1\over 2}\sin\theta e^{-i\phi }{f_0 f_2^*} \\
  {1\over 2}\sin\theta e^{i\phi }{f_0^* f_2} & \sin^2\frac{\theta}{2}|f_2|^2
 \end{array}\right).
\end{equation}
From Eqs. (2) and (4) one can show that $|f_0|=1$, thus if we define
$f=f_0^*f_2$ (when $f_0=1$, $f$ is just the transfer fidelity of an
excitation), then the reduced density matrix can be rewritten as
\begin{equation}
 \rho_{2}(t)=\left(\begin{array}{cc}
  1-\sin^2\frac{\theta}{2}|f|^2 & {1\over 2}\sin\theta e^{-i\phi }{f^*} \\
  {1\over 2}\sin\theta e^{i\phi }{f} & \sin^2\frac{\theta}{2}|f|^2
 \end{array}\right).
\end{equation}

   Eq. (6) enables one to compute the state transfer fidelity $F=\langle
\varphi_{in}|\rho_{2}(t)|\varphi_{in}\rangle$ as
\begin{equation}
\begin{split}
  F=&\cos^2\frac{\theta}{2}\left(1-|f|^2\sin ^2\frac{\theta}{2}+
     2|f|{\sin^2}\frac{\theta}{2}\cos\gamma\right) \\
    &+|f|^2\sin^4\frac{\theta}{2},
 \end{split}
\end{equation}
which yields
\begin{eqnarray}
 \bar F=\frac{1}{2}+\frac{|f|\cos\gamma}{3}+\frac{|f|^2}{6},
\end{eqnarray}
where $\gamma=\arg\{f\}$ denotes the argument of the complex number
$f$.

   From Eq. (8) it is easy to conclude that if we want to attain perfect
state transfer for all kinds of initial pure states (i.e., $\bar
F_{\rm{max}}=1$), we demand that $|f(t_c)|=1$ and
$\gamma(t_c)=2k\pi$ ($k\in \mathbb{Z}$), or equivalently,
${\rm{Re}}\{f(t_c)\}=1$ and ${\rm{Im}}\{f(t_c)\}=0$, where
${\rm{Re}}\{f(t_c)\}$ and ${\rm{Im}}\{f(t_c)\}$ represent the real
and imaginary part of $f$, respectively. So based on this
consideration, we only need to discuss effects of impurity on
dynamics of $f=f_0^* f_2$ in the following.

   To obtain the explicit expressions of $f_0=\langle\textbf{0}|
e^{-i\hat Ht}|\textbf{0}\rangle$ and $f_2=\langle 2|e^{-i\hat
Ht}|1\rangle$, one needs to obtain the eigenvalues as well as the
eigenvectors of the Hamiltonian $\hat H$. Since for the initial
state $|\psi(0)\rangle$ expressed in Eq. (1), its dynamics is
completely determined by the time evolution in the subspace spanned
by the site basis $\{|\textbf{0}\rangle, |1\rangle, |2\rangle\}$,
one can calculate the eigenvalues of the Hamiltonian $\hat H$ in
this subspace, which is $\epsilon_0={3\over 2}B$ and
${\epsilon_{1,2}}={1\over 2}(B\pm\sqrt 2J)$, with the corresponding
eigenstates given by
\begin{eqnarray}
  |\varphi_0\rangle=|00\rangle,\,
  |\varphi_{1,2}\rangle=\frac{1}{\sqrt 2}(|01\rangle\pm|10\rangle).
\end{eqnarray}

From the explicit expressions of the eigenstates given in Eq. (9),
one can obtain directly that $|00\rangle=|\varphi_0\rangle$ and
$|10\rangle=\frac{1}{\sqrt 2}(|\varphi_1\rangle-|\varphi_2\rangle)$.
Substituting these results into Eq. (4), one can obtain
\begin{eqnarray}
 f_0=e^{-i3Bt/2},\,f_2=-ie^{-iBt/2}\sin(\sqrt{2}Jt/2),
\end{eqnarray}
which yields
\begin{eqnarray}
 f=-ie^{iBt}\sin(\sqrt{2}Jt/2).
\end{eqnarray}

   Clearly, when $B=0$, the maximum value of the average fidelity
is $\bar F_{\rm{max}}=2/3$, which is attained at the critical time
$t_c=(2k+1)\pi/{\sqrt{2}J}~(k=0,1,2,\ldots)$, and equals to the best
possible score if Alice and Bob communicate with each other only via
a classical channel \cite{R35}. When $B\neq0$, however, one may
obtain $\bar F_{\rm{max}}=1$, i.e., all the purely input states
$|\varphi_{in}\rangle$ can be perfectly transferred. To show this is
true, we reconsider the critical time $t_c$ at which $\bar
F_{\rm{max}}=2/3$ in the absence of external magnetic field, from
which one can see that when $k\in even$, $\sin(\sqrt{2}Jt_c/2)=1$,
thus if one tunes the intensity of the external magnetic field to
$B_c=(4l+1)\pi/{2t_c}~(l=0,1,2,\ldots)$, then we have $f(t_c)=1$ and
$\cos\gamma(t_c)=1$, which gives rise to $\bar F_{\rm{max}}=1$.
Similarly, when $k\in odd$, we have $\sin(\sqrt{2}Jt_c/2)=-1$, thus
one needs to modulate the intensity of the external magnetic field
to $B_c=(4l+3)\pi/{2t_c}~(l=0,1,2,\ldots)$ in order to obtain the
maximum average fidelity $\bar F_{\rm{max}}=1$.

   When $B=0$, the average fidelity $\bar F$ can be maximized by
applying a local unitary operation to qubit 2 belonging to the
receiver Bob. Since at the critical time $t_c=(2k+1)\pi/{\sqrt{2}J}$
($k\in even$), we have $\sin(\sqrt{2}Jt_c/2)=1$, Bob can perform the
$S$ ($S=\sqrt Z$, with $Z=diag\{1,-1\}$ being the phase-flip gate)
operation to the qubit at his hands, which turns
$|0\rangle\mapsto|0\rangle$ and $|1\rangle\mapsto i|1\rangle$, and
thus gives rise to $f=\sin(\sqrt{2}Jt_c/2)=1$ and $\bar
F_{\rm{max}}=1$. Similarly, for the critical time
$t_c=(2k+1)\pi/{\sqrt{2}J}$ ($k\in odd$), since
$\sin(\sqrt{2}Jt_c/2)=-1$, Bob can perform the unitary operation
$U=SZ$ ($U|0\rangle=0$, $U|1\rangle=-i|1\rangle$) to the qubit at
his hands, which also yields $f=\sin(\sqrt{2}Jt_c/2)=1$ and $\bar
F_{\rm{max}}=1$.

   In the following we investigate efficiency of quantum state transfer
by using the three-spin XX chain as quantum channel, and we assume
the impurity spin is located at the central site of the system, then
the Hamiltonian can be written as
\begin{equation}
\begin{aligned}
 \hat H=&J(s_1^x S_2^x+s_1^y S_2^y+S_2^x s_3^x+S_2^y s_3^y) \\
        &+B(s_1^z+S_2^z+s_3^z).
 \end{aligned}
\end{equation}

   By using the same method, one can show that the explicit
expression of the average fidelity $\bar F$ for this system has the
same form as that expressed in Eq. (8), with however, the parameter
$f=f_0^* f_3$, with $f_0=\langle\textbf{0}|e^{-i\hat
Ht}|\textbf{0}\rangle$ and $f_3=\langle 3|e^{-i\hat Ht}|1\rangle$.
Thus in order to implement perfect state transfer for all kinds of
initial states $|\varphi_{in}\rangle$, we also require $f(t_c)=1$.

   For this system, its eigenvalues can be obtained explicitly as
$\epsilon_0=2B$, $\epsilon_1=B$, and $\epsilon_{2,3}=B\pm J$ in the
subspace spanned by $\{|\textbf{0}\rangle, |1\rangle, |2\rangle,
|3\rangle\}$, with the corresponding eigenstates given by
\begin{eqnarray}
 &|\varphi_0\rangle=|000\rangle,\,
   |\varphi_1\rangle=\frac{1}{\sqrt 2}(|001\rangle-|100\rangle ),\nonumber\\
 &|\varphi_{2,3}\rangle=\frac{1}{2}(|001\rangle\pm\sqrt 2|010\rangle+|100\rangle).
\end{eqnarray}

   From the above eigenstates, one can obtain directly that $|000\rangle=|
\varphi_0\rangle$ and $|100\rangle=\frac{1}{{2}}|\varphi_2\rangle+
\frac{1}{2}|\varphi_3\rangle-\frac{1}{\sqrt{2}}|\varphi_1\rangle $,
thus we have
\begin{eqnarray}
 f_0=e^{-i2Bt},\, f_3=-e^{-iBt}\sin^2(Jt/2),
\end{eqnarray}
which yields
\begin{eqnarray}
 f=-e^{iBt}\sin^2(Jt/2).
\end{eqnarray}

   In the absence of external magnetic field (i.e., $B=0$),
$\bar F_{\rm{max}}=1/2$, which corresponds to no-communication
between the sender Alice and the receiver Bob. However, at the
critical time $t_c=(2k+1)\pi/J~(k=0,1,2,\ldots)$, one can obtain the
maximum average fidelity $\bar F_{\rm{max}}=1$ when the intensity of
the external magnetic field $B_c=(2l+1)\pi/t_c~(l=0,1,2,\ldots)$.
Moreover, since $f=-\sin ^2(Jt/2)$ and $\gamma=\pi$ when $B=0$, one
can also maximize the average fidelity by performing the local
unitary operation $Z$ to the qubit 3 belonging to Bob, which gives
rise to $f=\sin ^2(Jt/2)$ and $\bar F_{\rm{max}}=1$.

   When the impurity spin locating at the edge site of the three-spin
XX chain, after a similar analysis as performed in the above
section, one can also show that perfect transfer of an arbitrary
one-qubit pure state is also possible by modulating the intensity of
the external magnetic field or performing relevant local unitary
operations to the spin belonging to the receiver Bob.

   In fact, even for the $N$-spin quantum channel with the presence of spin
impurity, one can still implement perfect transfer of an arbitrary
one-qubit state when the neighboring coupling strengths satisfying
${J_i}={J_{N-i}}=\lambda\sqrt{i(N-i)}$ $(i=1,2,\ldots,N-1)$, with
$\lambda$ being a scaling constant \cite{R11}. This is because for
the XX spin chain we always have the commutation relation $[\hat H,
S_{\rm{tot}}^z]=0$ (here $S_{\rm{tot}}^z=S_k^z+\sum_{i=1,i\neq k}^N
s_i^z$, i.e., the impurity spin locating at the $k$th site), which
ensures the initial state $|\varphi_{in}\rangle$ prepared in the
subspace $\{|\textbf{0}\rangle, |1\rangle, |2\rangle,\ldots,
|N\rangle\}$ will remains in it \cite{R23}, and the spin impurity
can only change the instant of time at which the state is perfectly
transferred.

\section{The quantum channel with magnetic impurity}

   From the above analysis one can see that the presence of spin
impurity does not prevent one from implementing perfect transfer of
an arbitrary one-qubit pure state from one end of the chain to
another. How does the other kinds of impurities or imperfections
affect efficiency of quantum state transfer across a spin chain? In
this section we will address this problem by introducing one
magnetic impurity \cite{R29} into the quantum channel, and discuss
dynamics of the average fidelity for quantum state transfer.

   We first discuss  quantum state transfer through a two-spin XX
chain in the presence of magnetic impurity, with the system
Hamiltonian given by
\begin{eqnarray}
\hat H=J(s_1^x s_2^x+s_1^y s_2^y)+Bs_1^z,
\end{eqnarray}
for which the eigenvalues can be obtained analytically as
$\epsilon_0={1\over 2}B$ and $\epsilon_{1,2}=\pm{1\over 2}\mu$ in
the subspace $\{|\textbf{0}\rangle, |1\rangle, |2\rangle\}$, with
$\mu=\sqrt{B^2+J^2}$, and the corresponding eigenstates are given by
\begin{eqnarray}
  &&|\varphi_0\rangle=|00\rangle,\nonumber\\
  &&|\varphi_{1,2}\rangle=\frac{1}{\sqrt {2\mu(\mu\pm B)}}
  [(B\pm\mu)|01\rangle+ J|10\rangle].
\end{eqnarray}

   For this kind of imperfect quantum spin channel, it can be shown
that the expression of the average fidelity $\bar F$ has the same
form as that expressed in Eq. (8). Moreover, from Eq. (17) one can
obtain directly that $|00\rangle=|\varphi_0\rangle$,
$|10\rangle=J|\varphi_1\rangle/\sqrt{2\mu(\mu+B)}+
 J|\varphi_2\rangle/\sqrt{2\mu(\mu-B)}$, which yields
\begin{eqnarray}
 f_0=e^{-iBt/2},\, f_2=-i{J\over\mu}\sin{\mu t\over2},
\end{eqnarray}
and
\begin{eqnarray}
 f=-ie^{iBt/2}{J\over\mu}\sin{\mu t\over2}.
\end{eqnarray}

   Since $J<\mu$ when $B\neq 0$, we have $|f|<1$ at any instant of
time, which implies that one cannot implement perfect transfer of an
arbitrary purely input state $|\varphi_{in}\rangle$ in the presence
of one magnetic impurity. This is different from that of the spin
impurity, which does not exclude the possibility of perfect state
transfer of $|\varphi_{in}\rangle$. Theoretically, for given $B$,
one can modulate the coupling strength $J$ of the neighboring spins
so that $J\gg B$, for which the parameter $f$ can be approximated by
$f\simeq-ie^{iBt/2}\sin{Jt\over2}$, thus when
$t_c=(4k+1)\pi/J,~B_c=(4l+1)\pi/t_c$ or
$t_c=(4k+3)\pi/J,~B_c=(4l+3)\pi/t_c~(k,l=0,1,2,\ldots)$, one can
obtain the maximum average fidelity $\bar F_{\rm{max}}\simeq 1$.
However, the realization of $J$ large enough might be a difficult
experimental task.

   Now we discuss average fidelity of quantum state transfer via the
three-spin channel, with the magnetic impurity locating at the
central site. The Hamiltonian of the system is give by
\begin{eqnarray}
 \hat H=J(s_1^x s_2^x+s_1^y s_2^y +s_2^x s_3^x+s_2^y s_3^y)+B s_2^z.
\end{eqnarray}

   In the subspace spanned by $\{|\textbf{0}\rangle,|1\rangle,
|2\rangle,|3\rangle\}$, the eigenvalues of the system can be
obtained as $\epsilon_{0,1}=\frac{1}{2}B$ and
$\epsilon_{2,3}=\pm{1\over 2}\nu$, with $\nu=\sqrt{B^2+2J^2}$ and
the corresponding eigenstates given by
\begin{eqnarray}
 &|\varphi_0\rangle=|000\rangle,\, |\varphi_1\rangle=\frac{1}{\sqrt 2}
  (|001\rangle-|100\rangle),\nonumber\\
 &|\varphi_{2,3}\rangle=\frac{1}{\sqrt{2\nu(\nu\mp B)}}[J|001\rangle-
  (B\mp\nu)|010\rangle+J|100\rangle],\nonumber\\
\end{eqnarray}
from which one can express $|000\rangle$ and $|100\rangle$ in terms
of the eigenstates as $|000\rangle=|\varphi_0\rangle$, and
$|100\rangle=J|\varphi_2\rangle/\sqrt{2\nu(\nu-B)}+
J|\varphi_3\rangle/\sqrt{2\nu(\nu+B)}-|\varphi_1\rangle/\sqrt{2}$.
Thus the parameters $f_0$, $f_3$ and $f=f_0^*f_3$ can be obtained
straightforwardly as
\begin{eqnarray}
 &&f_0=e^{-iBt/2},\nonumber\\
 &&f_3=\frac{J^2 e^{-i\nu t/2}}{2\nu(\nu-B)}+\frac{J^2 e^{i\nu
      t/2}}{2\nu(\nu+B)}-\frac{1}{2}e^{-iBt/2},
\end{eqnarray}
and
\begin{eqnarray}
 f=\frac{J^2 e^{i(B-\nu) t/2}}{2\nu(\nu-B)}+\frac{J^2 e^{i(B+\nu)
      t/2}}{2\nu(\nu+B)}-\frac{1}{2}.
\end{eqnarray}

   Similar to the two-spin channel, one still cannot obtain the
maximum average fidelity $\bar F_{\rm{max}}=1$ since $|f|<1$ and
$\gamma\neq 2k\pi~(k\in\mathbb{Z})$, i.e., the three-spin quantum
channel is also destroyed by the presence of the magnetic impurity.
Even when the coupling strength $J$ is strong enough, the average
fidelity attainable still cannot reach its maximum value 1 since
$f\simeq{1\over 2}[e^{iBt/2}\cos(Jt/\sqrt 2)-1]$ under the condition
of $J\gg B$.

   If the receiver Bob can perform a local unitary operation
$U$ to the spin at his hands, the average fidelity may be maximized
to a certain maximum but not unity. Since this requires $f(t_c)=1$
and $\gamma(t_c)=2k\pi~(k\in\mathbb{Z})$, the unitary operation $U$
must satisfying the following relations $U|0\rangle=|0\rangle$ and
$U|1\rangle=|f||1\rangle$, from which one can obtain
$U=diag\{1,e^{-i\vartheta}\}$, with
$\vartheta=\tan^{-1}[{\rm{Im}}(f)/{\rm{Re}}(f)]$. Using this method,
the average fidelity $\bar F$ can be greatly maximized. For example,
when $J=2\sqrt{2}B/3$, we have $\bar F_{\rm{max}}\simeq 0.9678$,
which is very close to its maximum value unity.

\section{The quantum channel with both spin and magnetic impurities}
   Now we investigate efficiency of quantum channel with both spin and
magnetic impurities. From the above two sections one can see that
the presence of spin impurity does not rule out the possibility of
perfect state transfer through an XX chain, while the magnetic
impurity may destroy the quantum channel and induce unavoidable loss
of quantum information during the transmission process, thus it is
natural to conjecture that under the influence of both spin and
magnetic impurities, the quantum channel may also be destroyed. To
show this is true, we consider the three-spin XX chain with both
spin and magnetic impurities locating at the central site, then the
Hamiltonian can be written as
\begin{eqnarray}
 \hat H=J(s_1^x S_2^x+s_1^y S_2^y +S_2^x s_3^x+S_2^ys_3^y)+BS_2^z.
\end{eqnarray}

   It can be shown that the explicit expression of the average
fidelity $\bar F$ has the same form as that expressed in Eq. (8),
with however, $f=f_0^*f_3$. Moreover, the eigenvalues of the system
can be calculated as $\epsilon_{0,1}=B$ and
$\epsilon_{2,3}=\frac{1}{2}(B\pm\xi)$ in the subspace spanned by the
site basis $\{|\textbf{0}\rangle, |1\rangle, |2\rangle,
|3\rangle\}$, with $\xi=\sqrt{B^2+4J^2}$, and the eigenvectors are
given by
\begin{eqnarray}
 &|\varphi_0\rangle=|000\rangle,\, |\varphi_1\rangle=\frac{1}{\sqrt 2}
 (|001\rangle-|100\rangle),\nonumber\\
 &|\varphi_{2,3}\rangle=\frac{1}{\sqrt{\xi(\xi\mp B)}}\left[J|001\rangle
                        -\frac{B\mp\xi}{\sqrt
                        2}|010\rangle+J|100\rangle\right].
\end{eqnarray}

   Thus $|000\rangle$ and $|100\rangle$ can be expressed in terms of
the eigenstates $|\varphi_i\rangle~(i=0,1,2,3)$ as
$|000\rangle=|\varphi_0\rangle$, and
$|100\rangle=J|\varphi_2\rangle/\sqrt{\xi(\xi-B)}+
J|\varphi_3\rangle/\sqrt{\xi(\xi+B)}-|\varphi_1\rangle/\sqrt{2}$,
which yields
\begin{eqnarray}
 &&f_0=e^{-iBt},\nonumber\\
 &&f_3=\frac{J^2 e^{-i(B+\xi)t/2}}{\xi(\xi-B)}+\frac{J^2 e^{-i(B-\xi)
     t/2}}{\xi(\xi+B)}-\frac{1}{2}e^{-iBt},~
\end{eqnarray}
and
\begin{eqnarray}
 f=\frac{J^2 e^{i(B-\xi)t/2}}{\xi(\xi-B)}+\frac{J^2 e^{i(B+\xi)
     t/2}}{\xi(\xi+B)}-\frac{1}{2}.
\end{eqnarray}

   Clearly, perfect transfer of all the purely input states
$|\varphi_{in}\rangle$ is also impossible since $|f|<1$ and
$\gamma\neq 2k\pi~(k\in \mathbb{Z})$. However, the average fidelity
$\bar F$ can also be maximized to a certain maximum value if Bob
performs a local unitary operation $U=diag\{1,e^{-i\delta}\}$ to the
spin at his hands, with
$\delta=\tan^{-1}[{\rm{Im}}(f)/{\rm{Re}}(f)]$. For example, when
$J=2B/3$, the average fidelity $\bar F$ can be adjusted to a certain
maximum value of about 0.9678 when the $U$ operation is performed.

\section{Conclusion}
   In conclusion, in this paper we have investigated effects of spin
and magnetic impurity on average fidelity of quantum state transfer
by using the XX spin chain as quantum channels. Our results revealed
that even in the presence of spin impurity, one can still implement
perfect transfer of an arbitrary one-qubit pure state by tuning the
strength of the external magnetic filed according to the instant of
time the receiver Bob decodes the information. One can also maximize
the average fidelity by performing relevant local unitary operations
at the spin belonging to Bob. When the magnetic impurity or both
spin and magnetic impurities are present, however, the quantum
channel is destroyed and thus one cannot obtain the maximum average
fidelity $\bar F_{\rm{max}}=1$, which implies that some information
is lost during the transmission process of the quantum states.
Though for some special cases (e.g., the three-spin quantum
channel), the average fidelity can be maximized to a certain maximum
value which is very close to unity by performing a proper local
unitary operation at the spin belonging to Bob, however, this
procedure does not work out for a general case.

\begin{acknowledgments}
   The author M.-L. Hu would like to acknowledge the financial support
by the Natural Science Foundation of Shaanxi Province under Grant
nos. 2010JM1011 and 2009JQ8006, the Specialized Research Program of
Education Department of Shaanxi Provincial Government under Grant
no. 2010JK843, and the Youth Foundation of Xi'an University of Posts
and Telecommunications under Grant no. ZL2010-32.
\end{acknowledgments}

\newcommand{\PRL}{Phys. Rev. Lett. }
\newcommand{\PRA}{Phys. Rev. A }
\newcommand{\PRB}{Phys. Rev. B }
\newcommand{\PLA}{Phys. Lett. A }

\end{document}